# Sub-single exciton optical gain threshold in colloidal semiconductor quantum wells with gradient alloy shelling

*Nima Taghipour,* [†] *Savas Delikanli,* [†,‡] *Sushant Shendre,* [‡] *Mustafa Sak,* [†] *Mingjie Li,* [§] *Furkan Isik,* [†] *Ibrahim Tanriover,* [†] *Burak Guzelturk,* [⊥] *Tze Chien Sum,* [§] *and Hilmi Volkan Demir* [†,‡,#]

[†] Department of Electrical and Electronics Engineering, Department of Physics, UNAM-Institute of Materials Science and Nanotechnology, Bilkent University, Ankara 06800, Turkey

[‡] Luminous! Centre of Excellence for Semiconductor Lighting and Displays, School of Electrical and Electronic Engineering, School of Physical and Mathematical Sciences, School of Materials Science and Engineering, Nanyang Technological University, Singapore 639798, Singapore

[§] School of Physical and Mathematical Sciences, Nanyang Technological University, Singapore 639798, Singapore

[⊥] Department of Materials Science and Engineering, Stanford University, CA 94025, USA

[#] Corresponding author: volkan@stanfordalumni.org

**Colloidal semiconductor quantum wells have emerged as a promising material platform for use in solution-processable light-generation including colloidal lasers. However, application relying on their optical gain suffer from a fundamental complication due to multi-excitonic nature of light amplification in common II-VI semiconductor nanocrystals. This undesirably increases the optical gain threshold and shortens the net gain lifetime because of fast nonradiative Auger decay. Here, we demonstrate sub-single exciton level of optical gain threshold in specially engineered CdSe/CdS@CdZnS core/crown@gradient-alloyed shell colloidal quantum wells. This sub-single exciton ensemble-averaged gain threshold of $\langle N_g \rangle \approx$ 0.80 (per particle) resulting from impeded Auger recombination along with a large absorption cross-section of quantum wells enables us to observe the amplified spontaneous emission starting at a low pump fluence of ~ 800 nJ cm$^{-2}$, at least three-folds better than the previously best reported values among all colloidal semiconductor nanocrystals. Moreover, long optical gain lifetimes of ~ 800 ps accompanied with modal gain coefficients of ~2,000 cm$^{-1}$ are achieved. Finally, using these gradient shelled quantum wells, we show a vertical cavity surface-emitting colloidal laser operating at an ultra-low lasing threshold of 7.5 µJ cm$^{-2}$. These results represent a significant step towards the realization of solution-processable electrically-driven colloidal lasers.**

Solution-processed semiconductor nanocrystals offer a versatile and promising gain medium for light amplification applications[1–4] owing to their broad spectral tunability[1,5,6], low-cost production and flexibility of using them in a broad range of matrices[1,4]. As a consequence, the utilization of colloidal semiconductor quantum dots (CQDs) as a gain medium has been experiencing a tremendous growth in the last few decades. In addition, alternative nanoemitters e.g., semiconductor nanorods[7] and perovskite nanocrystal[8,9], have recently emerged as an auspicious gain material for stimulated emission and lasing. Yet another promising candidate, colloidal semiconductor quantum wells (CQWs), the so-called nanoplatelets (NPLs)[10–12], have been shown to be excellent in optical gain applications thanks to their giant oscillator strength[13] and ultralarge modal gain coefficient[14]. Using different heterostructures of NPLs allows for ultra-low amplified spontaneous emission (ASE) thresholds[10,12]. However, their gain performance still suffers from nonradiative Auger recombination, wherein the released energy from recombination of the electron–hole is transferred to a third carrier. In common II-VI semiconductor nanocrystals, because of the non-unity degeneracy of the electron and hole states involved in emission, light amplification requires an ensemble-averaged number of excitons per nanocrystal greater than one ($\langle N \rangle > 1$), thereupon multi-excitonic Auger recombination strongly affects the dynamics of the carriers.

Previously, various efforts have been reported to tackle the issue of the fast Auger decay in nanocrystals including the smoothing of their confinement potential[15] and introducing the optical gain in single-exciton regime[2,7,16]. These methods have been shown to effectively enhance the optical gain performance of CQDs either by reducing influence or inactivating of the Auger process. On the other hand, in atomically flat CQWs, significant reduction in ASE and lasing threshold have previously been obtained in core/shell[12] and core/crown[10,11] heterostructures.

Nonetheless, the demonstrated approaches have not addressed the fundamental issue of the optical gain in CQWs where the multiexciton ($\langle N \rangle > 1$) nature of the light amplification results in fast decay of the carriers by a few hundred of picoseconds timescale[17,18] as a result of the Auger process. One way of inactivating of Auger recombination in CQWs is to employ the concept of single–exciton gain mechanism for optical amplification as formerly demonstrated in various heterostructures of the CQDs[2,7,16].

Here, we report optical gain in sub-single exciton regime ($\langle N \rangle < 1$) in quasi-type-II CdSe/CdS@Cd$_{1-x}$Zn$_x$S core/crown@gradient-alloyed shell (C/C@GS) CQWs tailored as a promising and novel solution for suppression of the Auger process with their smooth confinement potential, presenting an important step towards the evolution of semiconductor CQW lasers. We demonstrated an exceptionally low stimulated emission threshold of ~820 $nJ\ cm^{-2}$, corresponding to an average number of e–h pairs of 0.84 per NPL, which is also fully supported by nonlinear absorption measurements through ultrafast transient absorption spectroscopy. Sub-single exciton optical gain regime is also confirmed by linear dependence of the normalized absorption changes. The extremely large absorption cross-section (5.06 × 10$^{-13}$ cm$^2$) of our engineered NPLs, accompanied by an extremely large net modal gain coefficient of ~1,960 cm$^{-1}$, and a long net optical gain lifetime of ~830 ps result in such ultra-low optical gain thresholds and lead to record long stable ASE. Employing specifically engineered core/crown@gradient-alloyed shell heterostructures, we present a linearly polarized single-mode lasing from a vertical cavity surface-emitting laser enabling a record low lasing threshold of ~7.46 $\mu J\ cm^{-2}$.

For this study, we synthesized CdSe/CdS@Cd$_{1-x}$Zn$_x$S C/C@GS CQWs with a carefully controlled different number of Cd$_{1-x}$Zn$_x$S alloyed gradient shell monolayers (MLs), tuned from 2 to 6 MLs, grown on the seed of 4-ML CdSe/CdS core/crown NPLs by using colloidal atomic layer deposition (c-ALD) method[19]. Owing to the tuned gradient alloyed shell of the NPLs, electron wavefunction feels soft potential confinement in the vertical dimension and is largely relaxed into the shell due to the small conduction band-offset, while the hole wavefunction is mostly confined in the core region by virtue of the large valence band-offset[19]. This spatial separation between the electron and hole wavefunctions forms quasi-type-II electronic band alignment as previously shown in CdSe/CdS dot/rod heterostructure[7]. The schematic of C/C@GS NPL heterostructure and band-offsets of the heterostructure are given in Figures 1a and 1b, respectively. Transmission electron microscopy (TEM) images of the core, core/crown and core/crown@ shell NPLs are shown in Figures 1c-e, respectively. For the synthesis of C/C@GS hetero-NPLs, we started with 4-ML CdSe cores with lateral dimensions of 13 ± 3 nm, subsequently followed by 4-ML CdS crown (8 ± 1 nm) grown in the lateral direction, and then *via* using c- ALD method, a gradient alloyed shell of $Cd_{1-x}Zn_xS$ with 2, 3, 4, 5 and 6 MLs was grown on the core/crown NPLs. Further details of the syntheses procedures are given in Supporting Information (SI). The absorption and photoluminescence spectra of C/C@GS hetero-NPLs with 4 MLs of $Cd_{1-x}Zn_xS$ shell are depicted in Figure 1g. The inset of Figure 1g shows the absorption spectrum in a broader range of wavelength. The peaks appearing in the absorption spectrum at 630 and 572 nm are associated with the heavy- and light-hole excitonic transitions, respectively.

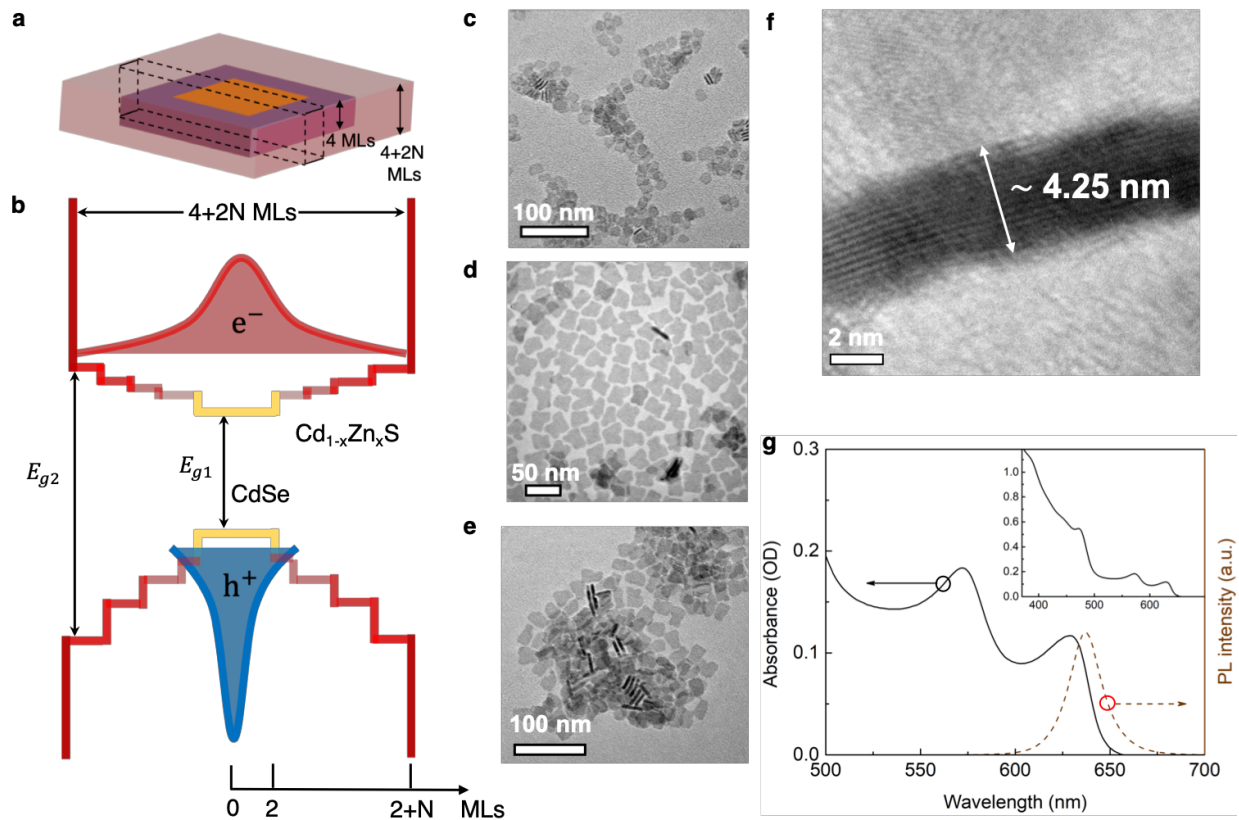

**Figure 1 | Structure and characterization C/C@GS of CdSe/CdS@Cd$_{1-x}$Zn$_x$S CQWs. a,** Schematic illustration of C/C@GS of CdSe/CdS@Cd$_{1-x}$Zn$_x$S NPLs synthesized by using interfacial grading of shell in the vertical dimension as detailed in the text. **b,** An approximate energy band diagram of the graded confinement potential in C/C@GS CQWs (based on band offset of bulk semiconductor[1]) along with a schematic representation of the spatial distribution of electron (e⁻) and hole (h⁺) wavefunctions. Here, $E_{g1} \approx 1.75$ eV, $E_{g2} \approx 2.7$ for 4+2N MLs, N= 4.[19] **(c-e)** TEM images of **c,** 4-ML core **d,** core/crown, **e,** C/C@GS CQWs. **f,** High-resolution TEM image of C/C@GS CQWs for 4+2N MLs, N= 4. The thickness equals ∼ 4.25 nm, which corresponds to 4 (core) + 2×4 (shell) MLs of core/shell heterostructure. **g,** Absorbance (black line) and PL (dashed red line) spectra of CdSe/CdS@Cd$_{1-x}$Zn$_x$S CQWs for 4+2N MLs, N=4. Inset shows the absorption spectrum of the same sample across a broader range of wavelength.

To analyze the optical gain performance of the CdSe/CdS@$Cd_{1-x}Zn_xS$ C/C@GS NPLs with different shell thicknesses, we performed femtosecond transient absorption (TA) measurements of NPLs in solution (hexane) with stirring (see Methods). TA measurements yield spectrally- and temporally-resolved pump-induced absorption changes ($\Delta\alpha$), which were used to calculate the nonlinear absorption spectra ($\alpha = \Delta\alpha + \alpha_o$) in the excited-state, where $\alpha_o$ is the absorption of the unexcited sample. Figure 2a presents $\alpha$ for the C/C@GS NPLs with 4 MLs of the shell coating (averaged over 1-5 ps pump-probe delay time) as a function of $\langle N_o \rangle$, where $N_o = f \times \sigma$ are the number of e–h pairs per NPL. Here, $f$ is the fluence and $\sigma$ is absorption cross-section of the NPL. The $\sigma$ is calculated based on the method that we discussed in our previous study[20] (see SI). As shown in Figure 2a, the band-edge absorption is gradually bleaching with increasing $\langle N_o \rangle$ and the transition to optical gain occurs when the absorption bleaching ($\Delta\alpha < 0$) is greater than the absorption of the unexcited sample ($\alpha_o$), therefore resulting in a negative net absorption ($\alpha < 0$). To determine the optical gain threshold ($\langle N_g \rangle$), we depict the normalized absorption bleaching ($-\Delta\alpha/\alpha_o$) as a function of $\langle N_o \rangle$ at the ASE peak position ($\lambda_{ASE} = 647$ nm). Then level of $\langle N_g \rangle = 0.78$ is quantified from the intersection of the measurement data (symbol) with horizontal dotted line at $|\Delta\alpha/\alpha_o| = 1$ (Fig. 2b). In addition, the growth of $|\Delta\alpha|$ follows a nearly linear function (red dashed-line) at initial values of $\langle N_o \rangle$ (Fig. 2b), this feature validates the sub-single exciton level of optical gain in these NPLs. Similar behavior was previously reported for type-II CQDs[16], where the optical gain occurs in the single-exciton regime.

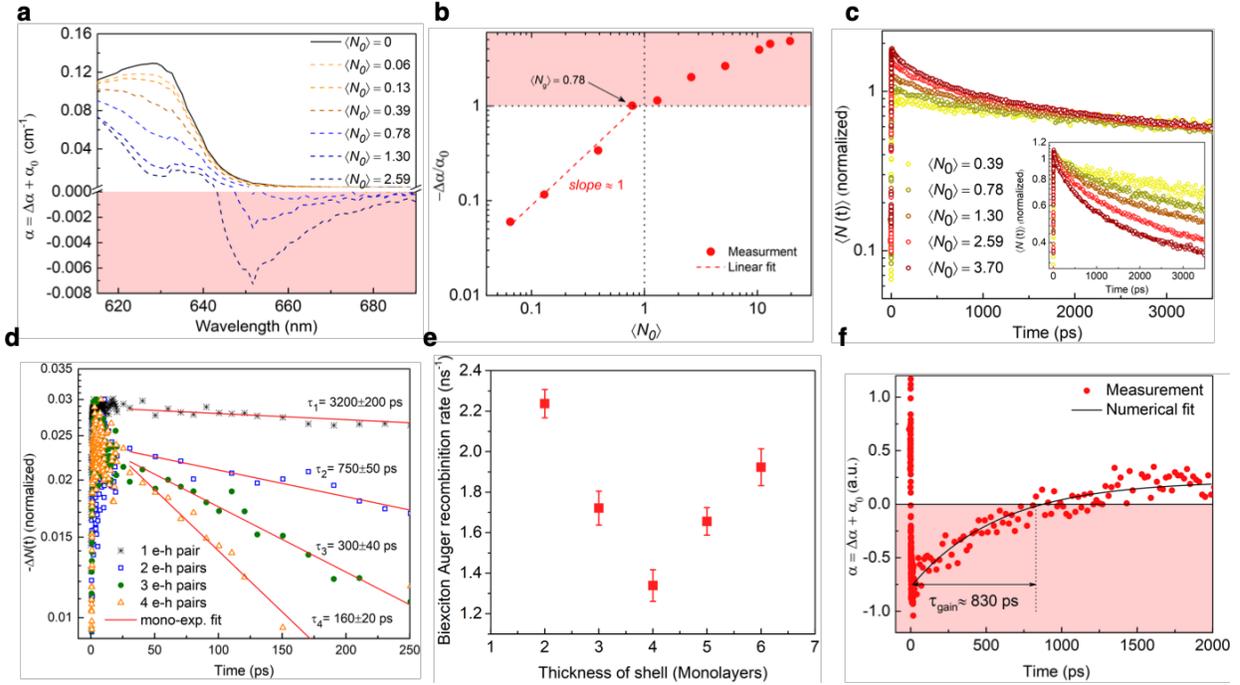

**Figure 2 | Optical gain characterizations of CdSe/CdS@Cd$_{1-x}$Zn$_x$S C/C@GS NPLs. a,** Nonlinear absorption ($\alpha$) of 4-ML shell NPLs as a function of $\langle N_o \rangle$ ( from 0 to 2.59, $N_o$ is the number of e–h pairs per NPL), obtained by averaging the recorded TA spectra at early time over 1-5 ps after pumping at 3.1 eV. Red-shaded region exhibits optical gain where $\alpha < 0$. **b,** Normalized absorption bleaching ($-\Delta\alpha/\alpha_o$) as a function of $\langle N_o \rangle$ at the ASE peak position ($\lambda_{ASE} = 647$ nm) with linear growth at low pump intensities. The red-shaded region corresponding to optical gain regime ($-\Delta\alpha/\alpha_o > 1$) implies a gain threshold of $\langle N_g \rangle = 0.78$. **c,** Dynamics of pump-dependent average population $\langle N(t) \rangle$ of 4 ML shell at ASE peak position, normalized to long term decay values. The inset shows the normalization to the initial values. **d,** Dynamics of 1, 2, 3 and 4 e–h pairs (symbols), obtained from TA spectra of 4 ML shell NPLs at the ASE peak position. The red solid line shows the fit to a single exponential decay. **e,** Biexciton Auger recombination rate as a function of the shell thickness. **f,** Nonlinear absorption ($\alpha$) as a function of time at the ASE peak position for $\langle N_o \rangle = 1.30$ (symbols). Shaded region corresponds to the net optical gain ($\alpha < 0$), indicating the net optical gain lifetime of $\tau_g \approx 830$ ps. The solid line is the numerical fitting.

To investigate the dynamics of the different e–h pair NPL states, we plotted the pump-dependent normalized average e–h pair populations $\langle N(t) \rangle$ per NPL at the peak position of stimulated emission ($\lambda_{ASE} = 647$ nm) (Fig. 2c). Here $\langle N(t) \rangle = \sum_N A_N e^{-t/\tau_N}$, where $A_N$ is the time-independent coefficient and $\tau_N$ is the lifetime of $N$-pair NPL state. Then, to analyze the multi-excitonic behavior of NPLs ($N \geq 2$), we employed a simple subtractive process to derive the single exponential decay dynamics as previously used for CQDs[21](See SI for details). The extracted single exponential dynamics are shown in Figure 2d, where $\tau_N$ corresponds to the lifetime of $N$ e–h pair NPL state. As seen in Figure 2d, by increasing the number of e–h pairs per NPL, the carrier decay kinetics becomes progressively faster, as expected for Auger recombination.

In the case of bulk semiconductor, the rate of Auger recombination ($\tau_N^{-1}$) follows $C_A(N/V_o)^2$ for $N \geq 2$ [21]. This expression results in a ratio of $\tau_4:\tau_3:\tau_2 = 0.25: 0.44: 1$ which is fully consistent with our experimental results, $\tau_4:\tau_3:\tau_2 = 0.22: 0.41: 1$, where the extracted lifetimes for 4, 3 and 2 e–h pairs correspond to $160 \pm 20$, $300 \pm 40$ and $750 \pm 50$ ps, respectively. Here, the single exciton lifetime (~3200 ps) is ~ 4-fold longer than the biexciton Auger lifetime. As expected, in the case of sub-single exciton gain regime, the inherent gain kinetics is characterized by the radiative single exciton lifetime, which is intrinsically much longer than the Auger decay lifetime.

Moreover, by performing the method as described above, we calculated the biexciton Auger recombination rate for 2, 3, 5, 6 MLs shell thicknesses. As can be seen in Figure 2e, the biexciton Auger decay rate exhibits a non-monotonic dependence on to the thickness of the shell. The origin of this behavior is the thickness-dependent oscillations in the overlap of the initial and final wavefunctions of the electron in the second exciton as previously shown in core/shell NPLs[22]. The temporal nonlinear absorption ($\alpha$) in the case of 4 ML thick shell at the ASE peak position for $\langle N_o \rangle = 1.30$ is presented in Figure 2f, in which the shaded region ($\alpha < 0$) corresponds to net optical gain. Here the gain lifetime was found to be ~830 ps from the intersection of the experimental data by the line of $\alpha = 0$. As expected, this value is very close to the biexciton Auger decay lifetime (750 ± 50 ps). It is worth mentioning that Auger recombination is suppressed in our engineered NPLs, while for simple core, core/crown and core/shell NPLs, the Auger decay lifetime was reported to be typically 150-500 ps[17,18,22]. The longer biexciton Auger life time in the gradient alloyed shell NPLs can be attributed to the significant reduction in the strength of the intraband transition as a result of the smooth potential confinement[23]. The fine grading of the confinement potential also contributes to the single-exciton gain regime in our gradient alloyed shell NPLs by suppressing Auger[15,23,24]. The abrupt interfaces were shown to exhibit higher Auger rates as a result of the larger overlap between the initial and final states of the intraband transition[15,23].

Next, we investigated the light amplification of CdSe/CdS@Cd$_{1-x}$Zn$_x$S C/C@GS CQWs, particularly for 4-MLs shell thickness by performing ASE and variable stripe length (VSL) measurements *via* stripe geometry excitation of the femtosecond mode-locked laser at 3.1 eV (see Methods). We prepared close-packed thin films of the engineered NPLs *via* spin-coating technique. The thickness of the ASE and VSL samples is $150 \pm 20$ nm. Pump-dependent PL spectra of the ASE measurement are shown in Figure 3a. As can be seen, by increasing the pump fluence,

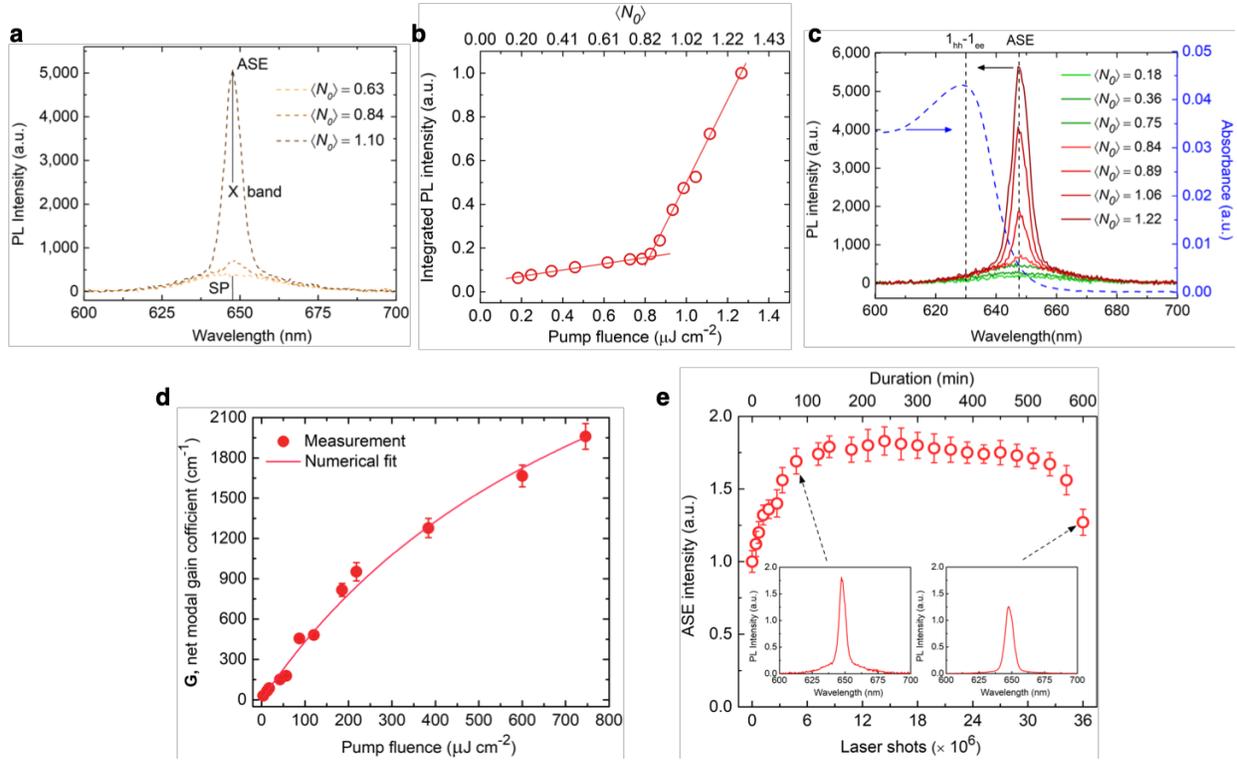

**Figure 3 | ASE, VSL and optical gain stability characterizations of C/C@GS of CdSe/CdS@Cd$_{1-x}$Zn$_x$S for 4-ML shell CQWs. a,** Pump-dependent ASE spectra at $\langle N_o \rangle = 0.63$, 0.84 and 1.10. The position of spontaneous emission (SP), ASE and X-band are marked by solid black line. **b,** Integrated PL intensity of ASE spectra as a function of the pump fluence and the corresponding average number of excitons per NPL ($\langle N_o \rangle$). **c,** Spectral analysis of ASE along with absorbance (dashed blue line) of the lowest excitonic energy state with respect to the $\langle N_o \rangle$. The position of $1_{hh}$-$1_e$ excitonic transition and ASE peak are marked by dark dashed lines. **d,** Pump-dependent net peak modal gain coefficient (symbols) at the ASE peak position ($\lambda_{ASE} = 647$ nm). The solid line is the numerical fitting to the experimental data, indicating a saturation fluence at ~920 $\mu$J cm$^{-2}$. The highest G reaches 1,960 ± 100 cm$^{-1}$ at 746 $\mu$J cm$^{-2}$ pump fluence. **e,** Pump-dependent ASE intensity as a function of the pump laser shot at $\langle N_o \rangle = 1.36$. The light amplification is highly stable in these engineered CQWs even for 9.5 h of continuous excitation.

a narrow emission feature develops at a position close to the spontaneous emission (SP) ($\lambda_{SP} = 646$ nm). The SP and ASE spectra exhibit full-width-at-half-maxima (FWHM) of ~28 and ~7 nm, respectively. This non-shifted ASE peak ($\lambda_{ASE} = 647$ nm), near the position of the single exciton band (X-band), exhibits a clear excitation threshold of ~820 $nJ\ cm^{-2}$, corresponding to $\langle N_o \rangle = 0.84$ above the threshold, where the integrated intensity of emission follows a super-linear function of the pump fluence. (Fig. 3b)

These important observations indicate that net optical gain regime occurs at the sub-single exciton level per particle in our CQWs having smooth confinement potential. In addition, the appearance of similar sharp emission is observed with increasing length of the stripe in VSL measurement while keeping the pump intensity fixed (see Methods). Also, the ASE measurements are fully consistent with our TA analysis in the solution form of these NPLs, which indicates an optical gain threshold of $\langle N_g \rangle = 0.78$. Here we reckon that the successful inactivation of Auger recombination allows to achieve the record lowest ASE threshold ($I_{ASE}$), which is several orders of magnitude lower than the CdS/ZnSe core/shell type-II CQDs[16] ($I_{ASE} \approx 2\ mJ\ cm^{-2}$), CdSe/ZnCdS core/shell type-I CQDs[2] ($I_{ASE} \approx 90\ \mu J\ cm^{-2}$) and charged CQDs[1] ($I_{ASE} \approx 6\ \mu J\ cm^{-2}$), in which the concepts of single- or sub-single exciton optical gain were reported. We attribute this superior optical gain performance of our engineered NPLs to their large absorption cross-section at the excitation wavelength ($\sigma = 5.06 \times 10^{-13}\ cm^2$ at 400 nm) owing to the core/crown heterostructure along with an efficient exciton funneling[10].

The lowest excitonic energy state (heavy-hole excitonic transition ($1_{hh}$-$1_e$)) is shown in Figure 3c, where the stimulated emission is largely Stokes-shifted with respected to the $1_{hh}$-$1_e$ excitonic feature. This major Stokes shift (~17 nm) significantly reduces the re-absorption of the emitted photon, a critical loss mechanism in optical gain, avoiding which leads to a remarkably reduced

ASE threshold and sub-single exciton optical gain regime. As seen in Figure 3c, the ASE peak emerges at the reddish tail of the absorption where the emitted photon at the ASE wavelength is absorbed insignificantly. This phenomenon was observed in gain media of colloidal CQDs[2], epitaxially-grown II-VI quantum well lasers[25] and organic-dyes[26] as reported previously.

The net modal gain coefficient (G) was also characterized *via* VSL measurement[10,27], which we performed at different pump intensities, by using the integrated intensity of the recorded PL spectra as a function of the stripe length ($l$) and employing the expression $I_{PL}(l) = A/G \, (e^{Gl} - 1)$[10,14], where A is a constant proportional to the spontaneous emission, the net modal gain coefficient (G) was evaluated (see SI). Figure 3d displays G with respect to the pump intensity; for instance, we achieved G = 460 ± 10 cm$^{-1}$ at an excitation fluence of 87 $\mu J \, cm^{-2}$. Here, by increasing the pump fluence, G rises first linearly at the low pump intensities, then followed by a gradual saturation behavior at elevated excitation fluences. The maximum achieved G is 1,960 ± 100 cm$^{-1}$ at 746 $\mu J \, cm^{-2}$. The obtained net modal gain is several-folds higher than the previously reported best performing heterostructures of semiconductor NCs, including core/shell CQDs[2] (G = 95 ± 10 cm$^{-1}$ at 120 $\mu J \, cm^{-2}$), core/crown NPLs[28] (G = 929 ± 23 cm$^{-1}$ at 580 $\mu J \, cm^{-2}$) and core/shell NPLs[10] (G = 600 ± 100 cm$^{-1}$ at 200 $\mu J \, cm^{-2}$). Our group also most recently reported an exceptionally large net modal gain in core NPLs of superior thin film quality as high as 6600 ± 350 cm$^{-1}$ [14].

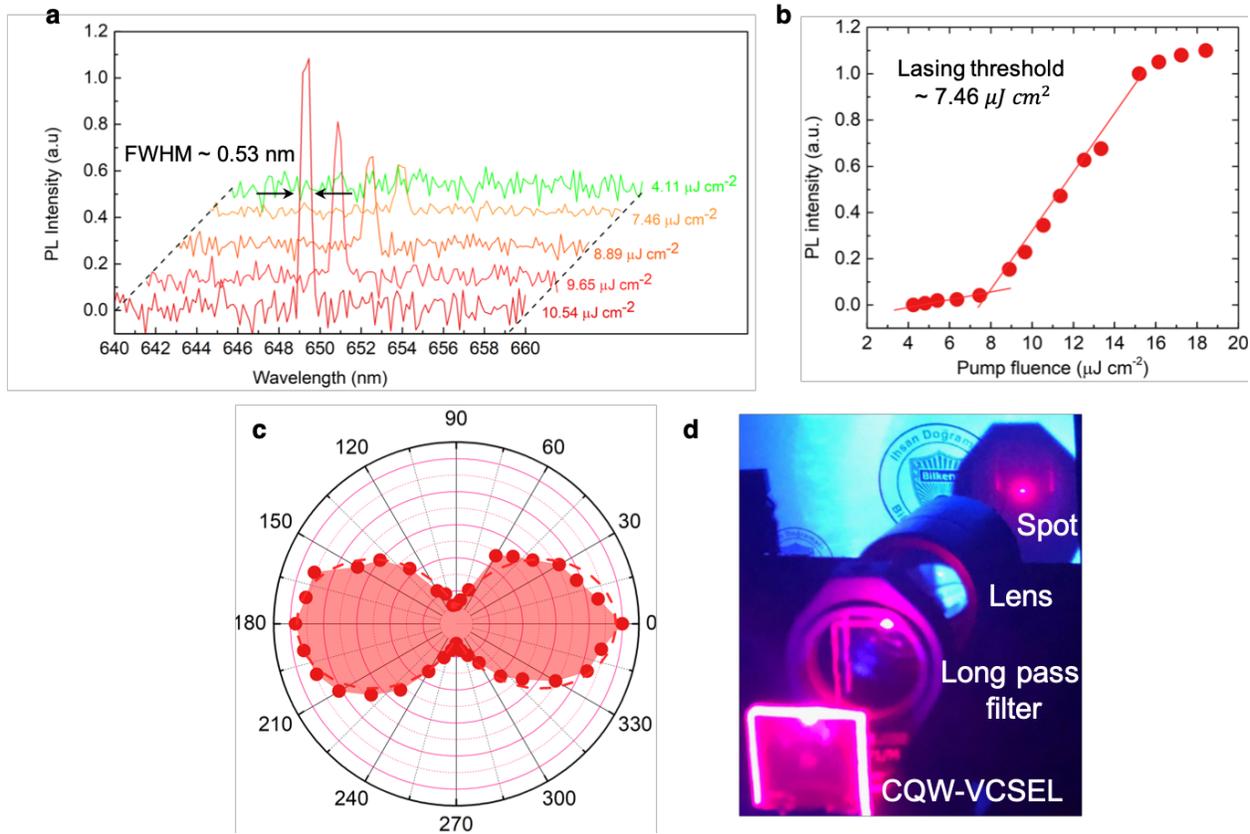

**Figure 4 | Lasing characterization of CQW-VCSEL. a,** PL spectra of CQWs-VCSEL as a function of the pump fluence. From the linewidth of the laser emission, the quality factor (Q) is calculated as ∼1,224. **b,** Emission intensity *versus* pump intensity (symbols). The red solid line indicating the lasing threshold of ∼7.46 $\mu J\ cm^{-2}$. **c,** Normalized emission spectrum of CQWs-VCSEL above the lasing threshold (∼12 $\mu J\ cm^{-2}$) as a function of the polarizer angle ($\theta$) and the detector at the front of the cavity. The dashed red line is the fitting of $\cos^2(\theta)$ curve. **d,** Photographical image of the CQWs-VCSEL above the lasing threshold indicating a well-defined spatial lasing spot on the screen.

To examine the stability of the light amplification in CdSe/CdS@Cd$_{1-x}$Zn$_x$S C/C@GS NPLs, we continuously pumped the ASE sample up to $36 \times 10^6$ laser shots (at 1 kHz repetition rate), which corresponds to overall 10 h of continuous excitation. The excitation intensity was ∼7 $\mu J\ cm^{-2}$, which is almost 10-fold larger than the ASE threshold. As shown in Figure 3e, the ASE intensity is stable up to ∼9.5 h, which is the record long stable ASE among all colloidal semiconductor NCs including CQDs[29] and NPLs[30], to the best of our knowledge. Therefore, our C/C@GS NPLs offer an effective architecture to substantially improve the optical gain stability issue of colloidal semiconductor NCs as a gain medium, owing to ultralow ASE threshold which is enabled by stimulated emission in the sub-single exciton regime and excellent passivation of the structure (top, bottom and lateral surfaces) with the crown and shell layers.

Finally, owing to highly efficient optical gain performance of our engineered CdSe/CdS@Cd$_{1-x}$Zn$_x$S C/C@GS CQWs, we incorporated these NPLs (with 4-ML shell thickness) as a gain medium into a vertical-cavity surface-emitting laser (VCSEL). The optical resonator was fabricated by using a pair of distributed Bragg reflectors (DBRs), the reflectance of each of which reaches ~96% at the emission wavelength of the NPLs while keeping lower than ~10% at the excitation wavelength (400 nm). Then, a close-packed solid film of these NPLs was sandwiched between the DBRs (see Methods). The emission spectrum of the CQW-VCSEL is depicted in Figure 4a at different pump fluences, demonstrating a single-mode lasing with a sharp laser output (FWHM ≈ 0.53 nm) at 649 nm corresponding to a $Q$-factor of ~1,224. The lasing output shows a clear excitation threshold of ~7.46 $\mu J\ cm^{-2}$ (Fig. 4b) with a well-defined spatial coherence profile (spot in Fig. 4d). This lasing threshold is the lowest among all previously reported colloidal semiconductor NC-based VCSELs, for which the next lowest reported lasing threshold is ~60 $\mu J\ cm^{-2}$.[2] As can be seen in Figure 4b, the pump-dependent emission displays an S-shaped curve (lasing intensity saturated at the elevated pump fluences > 15.2 $\mu J\ cm^{-2}$) which is one of the characteristic identifications of the lasing action. The ratio ($R$) of the polarization can be defined as $R = (I_\parallel - I_\perp)/(I_\parallel + I_\perp)$ [31], where $I_\parallel$ and $I_\perp$ is the intensity parallel and perpendicular to the optical axis, respectively. Figure 4c shows the lasing emission with respect to the polarizer angle, the dumbbell distribution of the measured data (symbols) indicating a linearly-polarized emission of our fabricated laser, which is well fitted by a quadratic cosine function ($\cos^2(\theta)$) where $R = 0.82$ (see SI).

In conclusion, the conducted study presents the first demonstration of light amplification in sub-single exciton gain regime of CQWs having smooth confinement potential, which avoids the complications associated with fast nonradiative multi-excitonic Auger process. This was confirmed by systematic nonlinear absorption analyses and linear dependence of normalized absorption changes below gain. This sub-single exciton ($\langle N_g \rangle \approx 0.8$) optical gain, accompanied with extremely large absorption cross-section, in CdSe/CdS@CdZnS core/crown@gradient-alloyed shell NPLs enables record low gain threshold among all colloidal semiconductor nanocrystals. The ultra-stable gain performance of these engineered NPLs addresses the stability problem of colloidal nanocrystals as a gain medium. Finally, the developed CQW–VCSEL exhibit a record low threshold of single-mode lasing. The realization of sub-single exciton regime optical amplification along with ultra-stable characteristics in these CQWs may enable lasing under continuous-wave excitation, possibly leading to electrically driven CQW-lasers.

## Methods

**Ultrafast transient absorption measurements.** TA spectroscopy was performed using a Helios™ setup (Ultrafast Systems LLC) and in transmission mode with chirp-correction. The white light continuum probe beam (in the range of 400-800 nm) was generated from a 3 mm sapphire crystal using 800 nm pulse from the regenerative amplifier. The 400-nm pump laser pulses were generated from a 1 kHz regenerative amplifier (Coherent Libra™). The beam from the regenerative amplifier has a center wavelength at 800 nm and a pulse width of around 150 fs and is seeded by a mode-locked Ti-sapphire oscillator (Coherent Vitesse, 80 MHz). The 400-nm pump laser was obtained by frequency-doubling the 800-nm fundamental regenerative amplifier output using a BBO crystal. The pump beam spot size was ~0.5 mm. The probe beam passing through the sample in cuvette was collected using a detector for UV–Vis (CMOS sensor). All measurements were performed at room temperature in solution (hexane) with stirring.

**ASE measurements.** As the pump source, we used a femtosecond mode-locked Ti: sapphire regenerative amplifier (Spectra Physics, Spitfire Pro) having a 120fs pulse width and a 1 kHz repetition rate, operating at the frequency-doubled output (400 nm by using a BBO crystal). A variable neutral density filter was employed to change the pump fluence on the sample. For the stripe geometry excitation, a cylindrical lens with a 10 cm focal length was used. Then, the photoluminescence signal was collected at the end of the stripe by a fiber coupled to a spectrometer (Maya 2000 Pro), where the collection was perpendicular to the excitation. For the ASE samples, we fabricated a thin film of CdSe/CdS@Cd$_{1-x}$Zn$_x$S C/C@GS CQWs by using a highly concentrated solution (40–50 mg mL$^{-1}$ of the CQWs in hexane) via spin-coating (at 1,000 rpm for 1 min) on quartz substrates.

**VSL measurements.** Sample preparation was very similar to ASE samples. The optical setup was slightly modified, where an adjustable slit was used to vary the length of stripe on the sample and also to minimize the effect of light diffraction, the sample was placed as close as possible to the slit (~4-5 mm). To ensure 1D amplifier assumption and to maximize the ASE intensity from the edge of sample, the length of the excitation was adjusted to be ~120 $\mu$m. The PL signal was collected at the edge of the sample.

**Fabrication of the CQW-VCSEL.** To fabricate highly reflective DBRs for CQW-VCSELs, eight pairs of $SiO_2/TiO_2$ (with the optical thickness of silica and titania chosen as quarter-wavelength emission of CQWs) were deposited on quartz substrate *via* sputtering. A narrow stripe of 120 $\mu$m-thick Kapton tape was placed at the edge of one of the DBRs. On the other side, using an epoxy, two DBRs were stick to each other by applying slight pressure, and then the tape was removed. This creates a finite wedge-shaped structure between two DBRs, which provides a built-in thickness variation in the resulting vertical cavity. Finally, a highly concentrated (50–60 mg mL$^{-1}$) CQW/hexane was poured and dried between two DBRs, together with which formed a cavity length variable Fabry-Perot resonator.

**Acknowledgment**

The authors gratefully acknowledge the financial support in part from Singapore National Research Foundation under the programs of NRF-NRFI2016-08, NRF-CRP14-2014-03 and the Science and Engineering Research Council, Agency for Science, Technology and Research (A*STAR) of Singapore and in part from TUBITAK 115E679. H.V.D. also acknowledges support from TUBA. T.C.S. and M. L. acknowledge the support from Singapore National Research Foundation through the Competitive Research Programme (CRP Award No. NRF-CRP14-2014-03).

**Author contributions**

N.T., S.D. and H.V.D. conceived the idea of this study and H.V.D supervised the research at all stages. S.D. and S.S. synthesized the engineered CQWs and characterized the optical properties of the samples in the solution form. N.T conducted the ASE and VSL measurements. N.T. and M.S. designed and fabricated the VCSEL along with the lasing measurements. M.L. and T.C.S conducted the pump-probe measurements. F.I. performed the ICP-MS measurements and contributed to the synthesis of CQWs. I.T. performed quantum mechanical calculation. N.T., S.D., B.G. and H.V.D analyzed and interpreted the data. N.T., S.D. and H.V.D wrote the manuscript with contributions from all authors.

Supporting Information

for

Sub-single exciton optical gain threshold in colloidal semiconductor quantum wells with gradient alloy shelling

*Nima Taghipour,* [†] *Savas Delikanli,* [†,‡] *Sushant Shendre,* [‡] *Mustafa Sak,* [†] *Mingjie Li,* [§] *Furkan Isik,* [†] *Ibrahim Tanriover,* [†] *Burak Guzelturk,* [⊥] *Tze Chien Sum,* [§] *and Hilmi Volkan Demir* [†,‡,#]

[†] Department of Electrical and Electronics Engineering, Department of Physics, UNAM-Institute of Materials Science and Nanotechnology, Bilkent University, Ankara 06800, Turkey

[‡] Luminous! Centre of Excellence for Semiconductor Lighting and Displays, School of Electrical and Electronic Engineering, School of Physical and Mathematical Sciences, School of Materials Science and Engineering, Nanyang Technological University, Singapore 639798, Singapore

[§] School of Physical and Mathematical Sciences, Nanyang Technological University, Singapore 639798, Singapore

[⊥] Department of Materials Science and Engineering, Stanford University, CA 94025, USA

## S1. Synthesis of core/crown@gradient-alloyed shell CdSe/CdS@Cd$_{1-x}$Zn$_x$S colloidal quantum wells

**Preparation of cadmium myristate:** Cadmium myristate was prepared according to a previously reported recipe[1]. 1.23 g of cadmium nitrate tetrahydrate was dissolved in 40 mL of methanol and 3.13 g of sodium myristate was dissolved in 250 mL of methanol. When both the powders were completely dissolved, the solutions were mixed and stirred vigorously for 1 h. Cadmium myristate was formed as a precipitate, which was then removed by centrifugation and washed by redispersing in methanol to remove any unreacted and/or excess precursors. After repeating the washing step for at least three times, the precipitated part was completely dried under vacuum overnight.

**Synthesis of the 4ML thick CdSe core NPLs:** The synthesis followed the recipe reported previously with slight modifications[1]. 170 mg of cadmium myristate, 12 mg of Se and 15 mL of ODE were loaded into a three-neck flask. After degassing the mixture for 1 h at room temperature, the solution was heated to 240 °C under argon atmosphere. When the solution turns bright orange (generally around 190-200 °C), 80 mg of cadmium acetate dihydrate was swiftly added to the reaction solution. After reaching 240 °C, the solution was cooled to room temperature and 0.5 mL of OA was injected. CdSe NPLs were precipitated by adding acetone and dispersed in hexane. Size-selective precipitation using centrifugation at different speeds was used if any additional sizes of NPLs were formed.

**Preparation of anisotropic growth solution for CdS crown:** A previously reported procedure was followed with slight modifications[1]. For the preparation of cadmium precursor, 480 mg of cadmium acetate dihydrate, 340 μL of OA, and 2 mL of ODE were loaded in a beaker. The solution was sonicated for 30 min at room temperature. Then, it was heated to 160 °C in ambient atmosphere under continuous stirring and alternating sonication until the formation of whitish color homogeneous gel. After the cadmium precursor was prepared, it was mixed with 3 mL of 0.1 M S-ODE stock solution and used for the CdS crown coating.

**Synthesis of CdSe/CdS core/crown NPLs:** A typical core-seeded synthesis method reported previously was used with slight modifications.[1] A portion of the 4 ML core synthesis in hexane and 15 mL of ODE were loaded in a three-neck flask. The solution was degassed at 100 °C for the complete removal of hexane. Then, the solution was heated to 240 °C under argon flow and a certain amount of anisotropic growth mixture for CdS crown was injected at the rate of 12 mL/h. After obtaining the desired crown size by adjusting the injection amount, the resulting mixture was further annealed at 240 °C for 5 min. After that, the solution was cooled down to room temperature and the core/crown NPLs were precipitated using ethanol. The NPLs were cleaned three times with ethanol and methanol to remove any traces of unreacted precursors, which was crucial for the shell growth step using c-ALD. Lastly, they were then dispersed in hexane to be used for the shell deposition.

**Synthesis of CdSe/CdS@$Cd_{1-x}Zn_xS$ core/crown@shell NPLs:** We used a modified procedure of our c-ALD recipe reported previously.[2,3] 1 mL of core/crown NPL seeds in hexane (having first absorption peak at 514 nm) were kept for use such that 100 μL of these NPLs dissolved in ~3 mL hexane had an optical density of ~2 at 370 nm. For cation precursors we used 0.4 M cadmium nitrate tetrahydrate (Cd-nitrate) and 0.4 M zinc nitrate hexahydrate (Zn-nitrate) solutions in NMF. For sulfur precursor we used 40-48 wt% solution of ammonium sulfide in water. Specifically, for the first sulfur shell growth, we added 40 μL of ammonium sulfide in 4 mL NMF and under vigorous stirring added 1 mL of CdSe/CdS core/crown seeds, which we had prepared separately. After 2 minutes of stirring when all the NPLs had entered the NMF phase from hexane, the reaction was stopped by quickly adding acetonitrile and excess toluene to precipitate the NPLs via centrifugation. This cleaning step was repeated once more by redispersing the NPLs in NMF and precipitating them using acetonitrile and excess toluene to remove any remaining sulfur precursor. Finally, the NPLs were dispersed in 4 mL of NMF for the next cation deposition step. Next, we added 1 mL of solution having a mixture of $X$% Cd-nitrate and (100-$X$)% Zn-nitrate in NMF by volume for the cation step. The $X$ (volume percentage of Cd precursor) value was varied as 50, 10, 5, 2, 1 and 1 for the first to the sixth $Cd_{1-x}Zn_xS$ shell layers. The reaction was allowed to continue by stirring for at least 45 min in ambient atmosphere and under room light, after which the NPLs were precipitated out by adding acetonitrile and excess toluene for centrifugation. The cleaning step similar to the previous one was repeated once more to remove all excess precursors. The growth cycle of sulfur and cation precursors were further continued (now only in NMF) to increase the number of shells as required while gradually increasing the Zn content with each cycle as indicated above. Each cycle added 1 monolayer of $Cd_{1-x}Zn_xS$ shell on top of the previous. At the end of each cycle a small amount of the NPLs were precipitated from NMF and dispersed in hexane by adding oleylamine for further use.

## S2. Quantum mechanical calculation of CdSe/CdS@Cd$_{1-x}$Zn$_x$S core/crown@gradient-alloyed shell colloidal quantum wells

**Calculations of electron and hole wavefunctions:** To calculate the electron and hole wavefunctions for the first excited state, we solved time–independent Schrödinger equation (TISE) for the potential profile given in Figure S1:

$$\left[\frac{\hbar^2 d^2}{2m_d z^2} + V(z)\right]\Psi(z) = E_z \Psi(z) \quad (S1)$$

where $m_z$ is the effective mass of electron (hole), $V(z)$ is the potential arising from conduction (valence) band offsets, and $\Psi(z)$ is the wave function of electron (hole). We assumed the even parity for the symmetry of first excited states and solved the TISE for $z > 0$. Then, because of the symmetry, we reflected the obtained solution to $z < 0$.

To solve the problem, we divided the right half of the potential well into five regions, as indicated in Figure S1, and write the general solutions for each region:

Region I, by assuming even symmetry:

$$\Psi(z) = A\cos(kz) \quad (S2)$$

where $k = \sqrt{\frac{2m_1 E}{\hbar^2}}$.

For Regions II, III, IV and V:

$$\Psi(z) = C_{2,3,4,5} e^{K_{2,3,4,5} z} + D_{2,3,4,5} e^{-K_{2,3,4,5} z} \quad (S3)$$

where $K_{2,3,4,5} = \sqrt{\frac{2m_{2,3,4,5}(V_{1,2,3,4} - E)}{\hbar^2}}$.

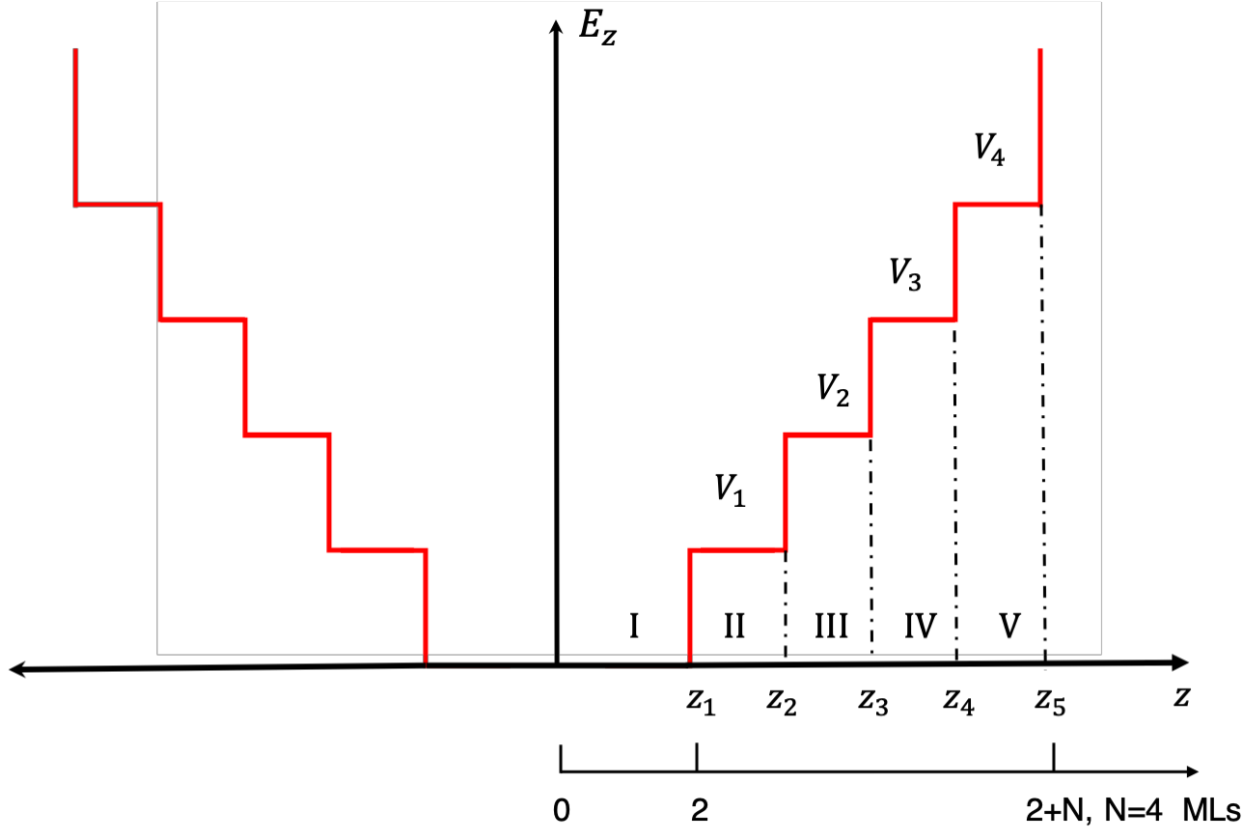

**Figure S1. Potential diagram of CdSe core coated with N=4-MLs Cd$_{1-x}$Zn$_x$S gradient alloyed-shell.** Here, x represents the amount of Zn in the shell region, where x is 0.02%, 13.5%, 24.6% and 53.2% for 1$^{st}$, 2$^{nd}$, 3$^{rd}$ and 4$^{th}$ MLs of the shell coating, respectively. $V_1$ is the potential barrier offset between 4-ML CdSe core and CdS shell. $V_{2-4}$ are the potential barrier offsets for the electron (or hole) wavefunctions between CdSe core and Cd$_{1-x}$Zn$_x$S shell for different amount of x.

Then, we solved the equations for electron (hole) by using the following boundary conditions:

$$\Psi(z^-_{1,2,3,4}) = \Psi(z^+_{1,2,3,4}) \qquad (S4)$$

$$\frac{d\Psi(z)}{dz}\Big|_{z^-_{1,2,3,4}} = \frac{d\Psi(z)}{dz}\Big|_{z^+_{1,2,3,4}} \qquad (S5)$$

$$\Psi(z_5) = 0 \qquad (S6)$$

To do so, "A" assumed to be 1 for finding the general solution, and then it was obtained by applying the normalization condition:

$$\int_{-\infty}^{+\infty}|\Psi|^2 dz = 1 \qquad (S7)$$

In our calculations, for the value of effective masses and band offsets, we used the data that are given in Ref. S3.

**Solutions;**

**At the interface between Region I-II ($z = z_1$):**

From Equations S2 and S3 and by applying the boundary conditions of S4 and S5:

$$C_2 = \frac{K_2 \cos(kz_1) - k \sin(kz_1)}{2K_2 e^{K_2 z_1}} \text{ and } D_2 = \frac{K_2 \cos(kz_1) + k \sin(kz_1)}{2K_2 e^{-K_2 z_1}}$$

**In region II-III, III-IV and IV-V interfaces ($z = z_{2,3,4}$)**

From equation 3 and applying the boundary conditions of S4 and S5:

$$C_{3,4,5} = \frac{C_{2,3,4}(K_{3,4,5}+K_{2,3,4})e^{K_{2,3,4,5} z_{2,3,4}} + D_{2,3,4}(K_{3,4,5}-K_{2,3,4})e^{-K_{2,3,4,5} z_{2,3,4}}}{2K_{3,4,5} e^{K_{3,4,5} z_{2,3,4}}} \qquad (S8)$$

and

$$D_{3,4,5} = \frac{C_{2,3,4}(K_{3,4,5}-K_{2,3,4})e^{K_{2,3,4,5} z_{2,3,4}} + D_{2,3,4}(K_{3,4,5}+K_{2,3,4})e^{-K_{2,3,4,5} z_{2,3,4}}}{2K_{3,4,5} e^{-K_{3,4,5} z_{2,3,4}}} \qquad (S9)$$

Finally, by applying the boundary condition S6, we calculated the first excited energy state ($E$) and corresponding electron (hole) wavefunctions, $\Psi(z)$. The $[\Psi(z)]^2$ is depicted as function of z in Figure S2.

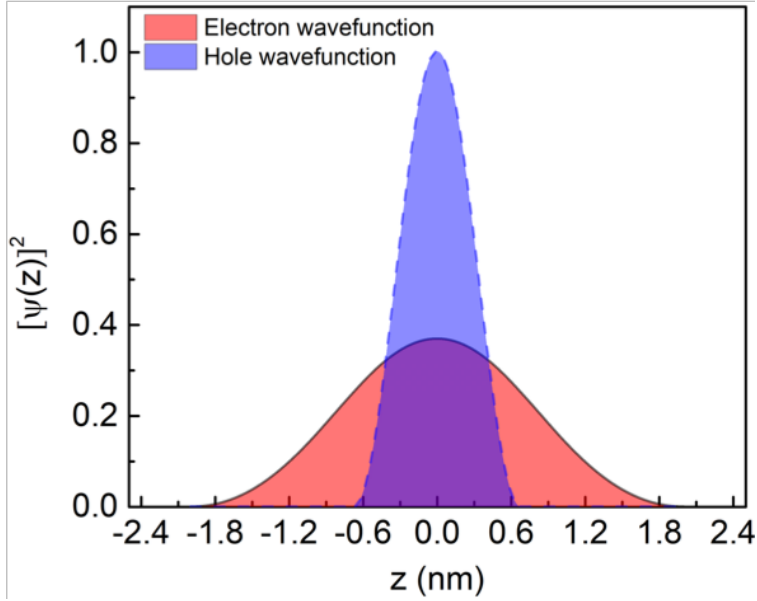

**Figure S2. Distribution of electron and hole wavefunctions.** Calculated the square of electron (red) and hole (blue) wavefunctions for CdSe/Cd$_{1-x}$Zn$_x$S core/alloyed-shell of 4+2N MLs, with N=4 MLs (shell thickness).

As can be seen in Figure S2, in our engineered heterostructure of CQWs, the hole wavefunction is mostly confined in the core region due to large band offset (0.52 eV) and heavy effective mass (0.89 m$_o$) from $-0.6$ to $+0.6$ nm in Figure S2. However, the electron wavefunction relaxes to the shell region, because of the weak potential barrier of the conduction band (~0.1eV) and light electron effective mass (0.13 m$_o$). Consequently, this partial separation of electron and hole wavefunctions results in quasi-type-II band alignment.

## S3. Calculation of average number of excitons per CQW

**Absorption cross-section calculations of NPLs:** First, inductively coupled plasma-mass spectroscopy (ICP-MS) measurement was used to calculate the concentration of NPLs. For the ICP-MS measurements, a solution of 1 mL of NPL sample was carefully dried. Then, the residual was dissolved in 5 mL nitric acid solution. The processed sample was placed into a volumetric flask for the ICP-MS measurement. The cadmium and zinc molar concentrations were obtained using an Agilent 725 ICP-MS system.

The size of the CdSe NPLs were obtained from transmission electron microscopy (TEM) images. NPL concentration ($C_{NPL}$) was calculated using $C_{NPL} = C \times (Cd+Zn) \times V_{unit}/(4V_{NPL})$, where $V_{unit}$ is the volume of the CdSe unit cell, $V_{NPL}$ is the physical volume of CdSe NPL and $C \times (Cd+Zn)$ is the total concentration of the Cd and Zn[4]. Herein, we make an assumption that the density of the NPLs is the same as its bulk material. After the calculation of NPL concentration, per-particle absorption cross-section ($\sigma$) of CdSe/CdS@CdZnS core/crown/shell NPLs was calculated by employing the equation of $\sigma = 2303A/(C_{NPL} \times N_A \times L)$, where A is the absorbance, $N_A$ is the Avogadro's number, and L is the optical path length.

**Calculation of Zn concentrations in the CdSe/CdS@CdZnS core/crown/shell NPLs:** Zn concentrations in the NPLs were acquired by using ICP-MS measurements. First, Zn concentration of all samples was calculated through ICP-MS measurements. Then, the Zn concentration at each shell layer was calculated by knowing the relative Zn concentration of each sample acquired through consecutive c-ALD procedures and considering the fact that each monolayer has the same volume as a result of their 2D planar geometry.

The calculated Zn concentration in each layer of shells is given in Table S1.

**Table S1.** Calculated the Zn concentration of in CdSe/CdS@Cd$_{1-x}$Zn$_x$S core/crown@gradient-alloyed shell NPLs.

| Shell number | Zn concentration (%) |
|---|---|
| 1 | 0.02 |
| 2 | 13.50 |
| 3 | 24.60 |
| 4 | 53.20 |
| 5 | 81.60 |
| 6 | 86.50 |

## S4. Biexciton Auger recombination lifetime calculation.

By using the method that we mentioned in the main text, we employed a simple subtractive process to derive single exponential decay dynamics as previously used for CQDs[5]. The biexciton Auger recombination rate as a function of the shell layers is shown in Figure 2e.

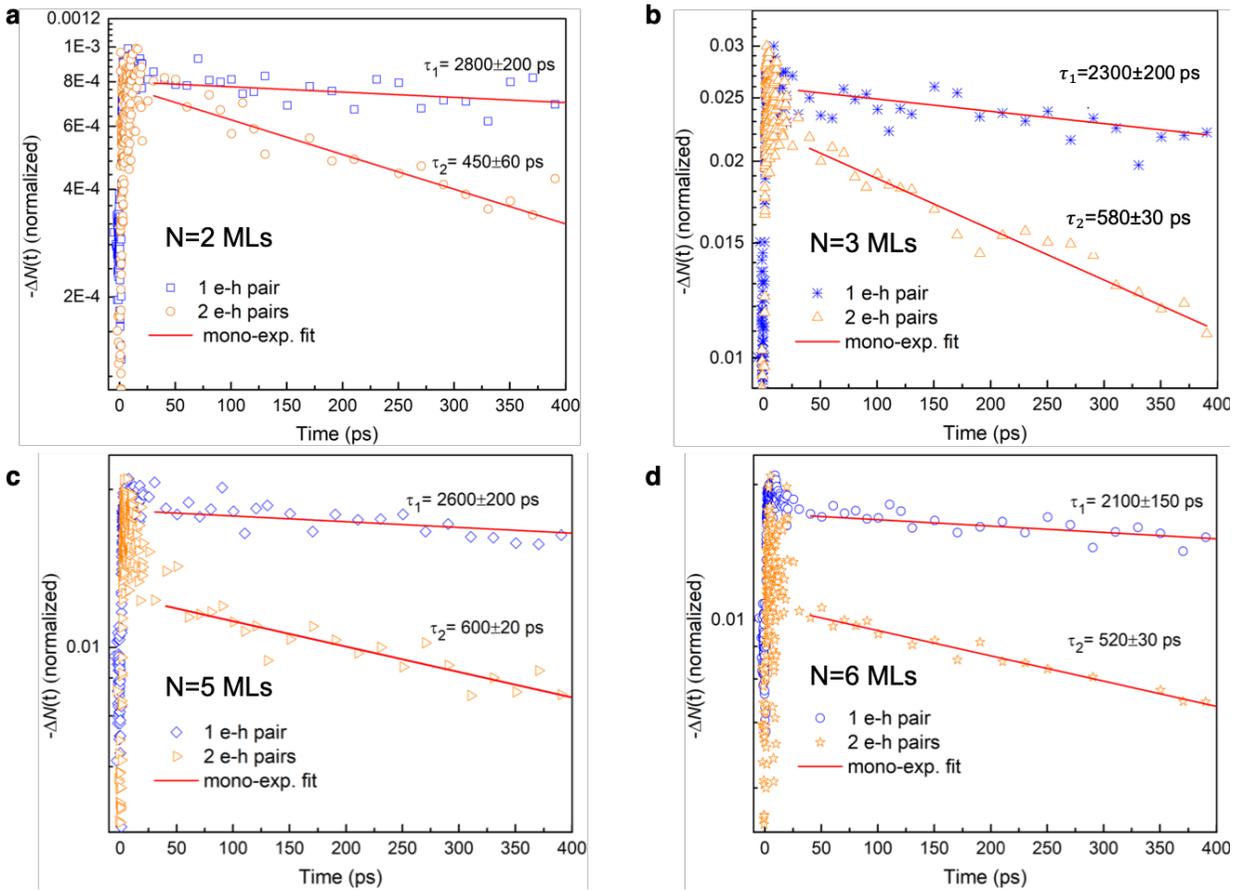

**Figure S2. Calculated biexciton Auger recombination lifetime for** CdSe/Cd$_{1-x}$Zn$_x$S core/alloyed-shell CQWs with different shell thicknesses. **a,** N=2, **b,** N=3, **c,** N=5 and **d,** N=6 MLs.

## S5. Amplified spontaneous emission measurement

We examined amplified spontaneous emission (ASE) of different samples in the same setup. To do so, we prepared the thin films of CdSe/CdS@Cd$_{1-x}$Zn$_x$S C/C@GS CQWs, by applying the similar recipe for all samples.

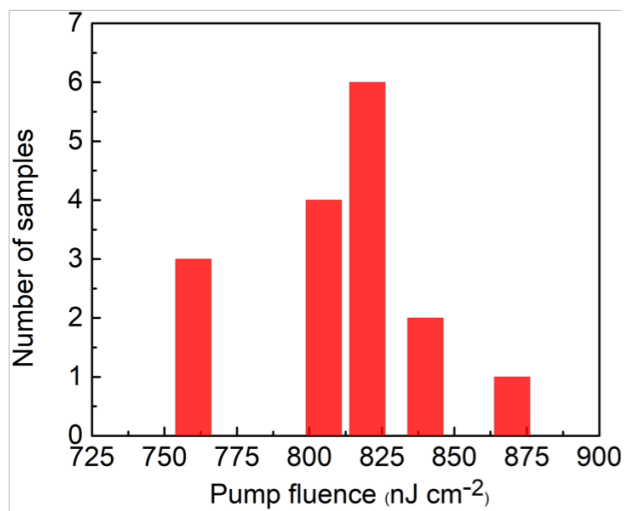

**Figure S3. Amplified spontaneous emission characterizations.** Histogram of ASE thresholds for different thin films of CdSe/CdS@Cd$_{1-x}$Zn$_x$S CQWs for 4-ML shell.

## S6. Variable stripe length measurement

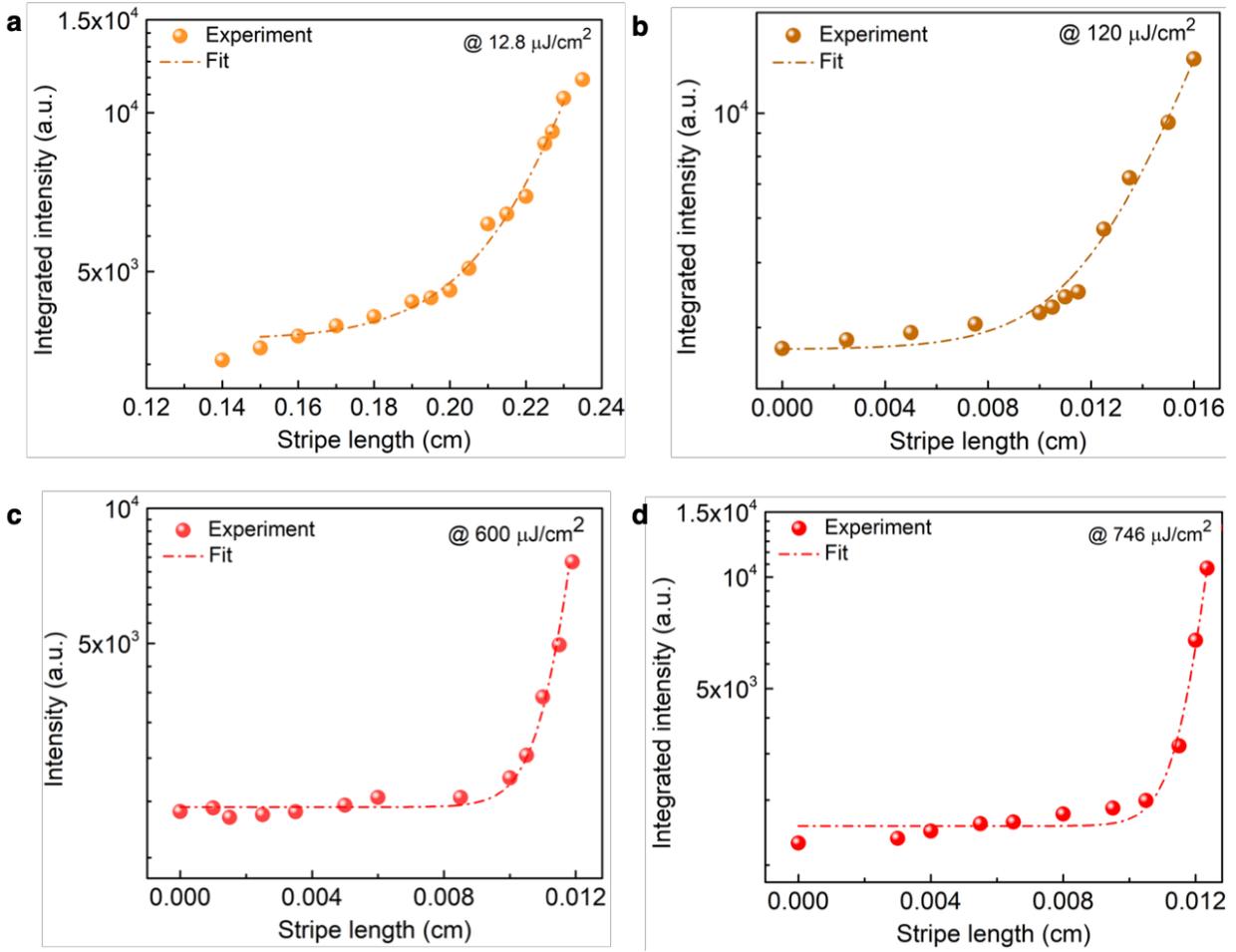

**Figure S4. Variable stripe length measurements of CdSe/CdS@Cd$_{1-x}$Zn$_x$S CQWs for 4-ML shell at different pump fluences.** a) 12.80, b) 120, c) 384, and d) 746 $\mu$J cm$^{-2}$ for which the modal gain coefficient corresponds to 60, 480, 1650 and 1960 cm$^{-1}$, respectively.

## S7. Colloidal quantum well- vertical- cavity surface-emitting laser.

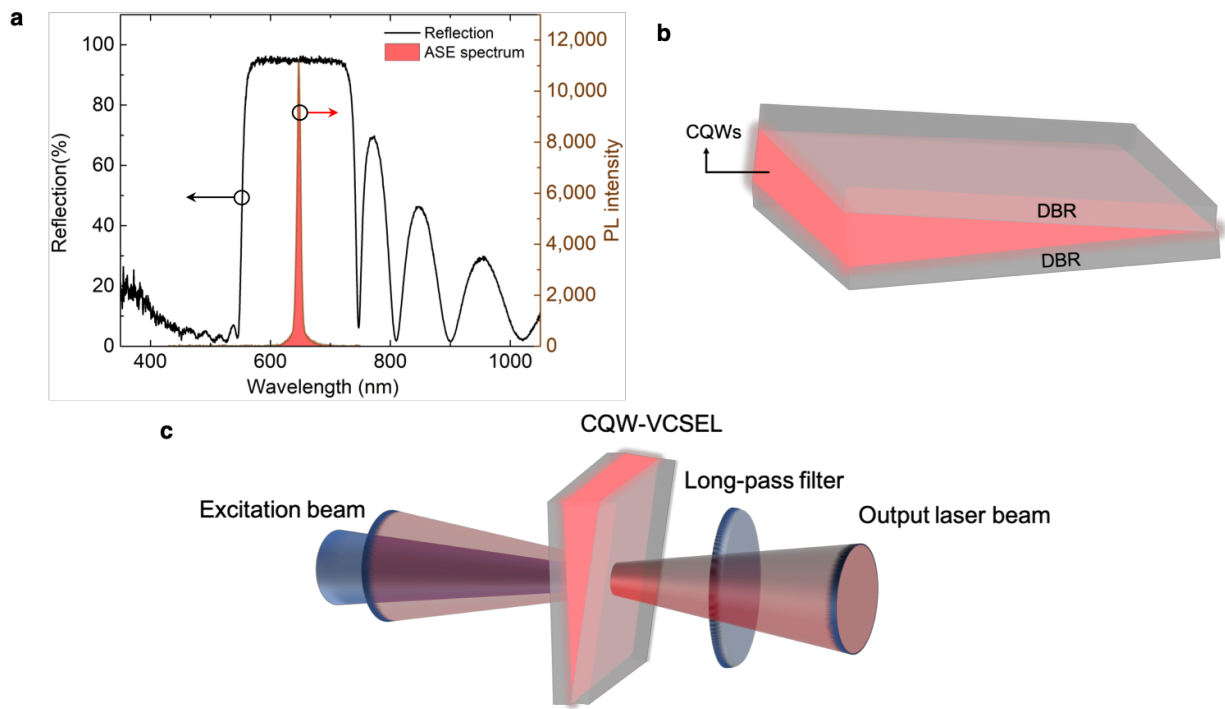

**Figure S5. CQW-VCSEL.** a, Reflection spectra of the DBRs and ASE spectrum of CdSe/CdS@$Cd_{1-x}Zn_xS$ CQWs with 4-ML shell. b, Schematic illustration of wedge CQW-VCSEL consists of two DBRs having a reflection of ~96% where a close-packed solid film of CQWs is sandwiched between them. c, Schematic of vertically-pumped CQW-VCSEL wedge which provides a variable cavity length. To eliminate any residual excitation beam, we used a long-pass filter. The output laser beam was collected with a spectrometer.

**Polarization measurements of output laser beam:** To characterize the polarization state of the laser emission, a linear polarizer was placed between the cavity output and spectrometer. By varying the angle of the polarizer, the signal was collected at the spectrometer. The luminescence intensity of the cavity is depicted as function of the detection angle in Figure S6, where the polarization ratio was found to be 9.09.

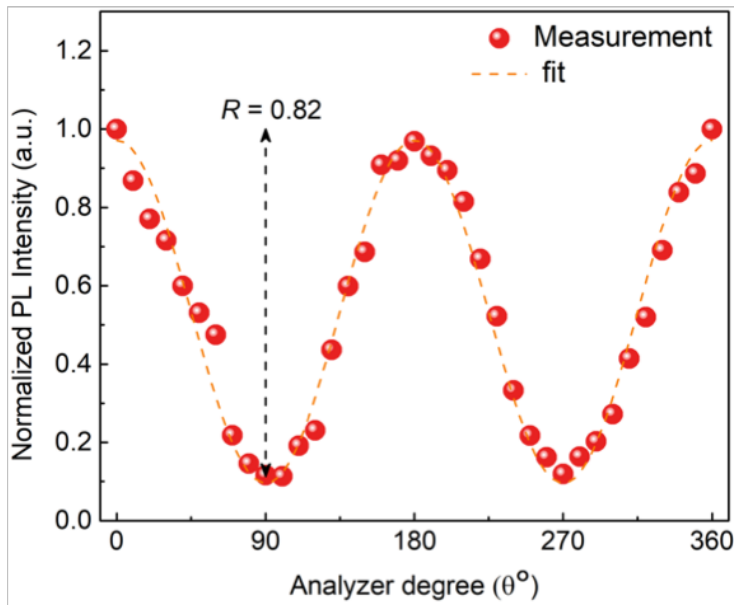

**Figure S6.** PL intensity of the collected signal *versus* detection angle. The polarization factor was calculated using experssion of $R = (I_\parallel - I_\perp)/(I_\parallel + I_\perp)$, which leads to 0.82. The experimental data is well fit with a $Cos^2$ function.

**SI Refrences**